\documentclass[12pt]{article}

\usepackage{epsfig}
\usepackage{amssymb}
\usepackage{amsfonts}

\usepackage{color}

\def\be{\begin{equation}}
\def\ee{\end{equation}}
\def\ba{\begin{array}{c}}
\def\ea{\end{array}}

\newcommand{\bea}{\begin{eqnarray}}
\newcommand{\eea}{\end{eqnarray}}

\newcommand{\bbr}{\br\!\br}
\newcommand{\kkt}{\kt\!\kt}

\newcommand{\kt}{\rangle}
\newcommand{\br}{\langle}

\begin{document}

\titlepage


 \begin{center}{\Large \bf

{Quantum singularities in a solvable toy model}

  }\end{center}


 \begin{center}

\vspace{8mm}

  {\bf Miloslav Znojil} $^{1,2,3}$

\end{center}

\vspace{8mm}
%
%
%
%
%
%
%



{$^1$
Department of Theoretical Physics, Nuclear Physics
Institute of The Czech Academy of Sciences,  \v{R}e\v{z}, Czech
Republic}

 {$^2$
 Department of Physics, Faculty of Science of the
 University of Hradec Kr\'{a}lov\'{e}, 
 Hradec Kr\'{a}lov\'{e},  Czech Republic}

{$^3$
Institute of System Science, Durban University of
Technology, Durban,
 South Africa}




\section*{Abstract}

Via elementary examples it is demonstrated that
the singularities of classical physics (sampled by the Big Bang
in cosmology) need not necessarily get smeared out
after quantization. It is proposed that the
role of quantum singularities
can be played by the
so called Kato's
exceptional-point spectral degeneracies.


\section{Motivation}

One of the decisive obstacles causing the current non-existence of a
fully consistent quantum theory of gravity \cite{Rovelli,Thiemann}
may be seen in the
comparatively less
developed nature of the quantum-theoretical component of the
prospective synthesis.
As a consequence, the
existence of various types of singularities
in the classical Einstein's theory of gravity
is widely believed to be incompatible with the intuitive perception
of the process of quantization after which one expects
that the singularities get ``smeared out''.

In our present brief communication we intend to oppose the
latter, widely
accepted assumption that
the classical singularities
must necessarily get regularized and
disappear after quantization.
Via a few schematic elementary examples we will
outline and support the possibility of survival of
at least some of the singularities
even after an amendment of their description using quantum theory.

\section{Mathematical aspects of the singularities}

From the point of view of physics
our present study has been inspired
by the cosmological
concept of Big Bang
emerging, very naturally,
in the framework of classical physics.
In contrast, in
the quantized descendants of the theory
(and, mainly,
in the most ambitious loop quantum gravity (LQG) approach),
this concept
is currently being replaced
by
its non-singular, regularized
``Big Bounce'' alternative \cite{Bounce,piech}.

\subsection{Regularizations and the concept of Big Bounce in quantum cosmology}

In the language of mathematics,
even on a very elementary level,
the latter change of paradigm
does not look too surprising.
Even when one considers just
a quantum-mechanical system represented
in an elementary $N-$dimensional Hilbert space
${\cal H}_{(physical)}$
and characterized by the Hermitian and
time-dependent $N$ by $N$
matrix observables
  \be
  R_{(Hermitian)}^{(N)}(t)=\left [R_{(Hermitian)}^{(N)}(t)\right ]^\dagger
  \label{txtb}
  \ee
it is immediate to conclude that
all of the classical
singularities (and, in particular,
the Big-Bang-like
coincidence of eigenvalues)
disappear after the quantization.

One of the underlying tacit assumptions
is that the physical
Hilbert space of states ${\cal H}_{(physical)}$
is unique, with an inner product which
remains time-independent.
In such a setting
the authors of the traditional textbooks on quantum mechanics
(cf., e.g., \cite{Messiah})
emphasize that
the eigenvalues  of $R_{(Hermitian)}^{(N)}(t)$ would,
in the absence of an additional {\it ad hoc} symmetry,
``never cross''.
Thus, even though
the classical limit
of the related observable may be singular (say, at $t=0$),
its quantum (i.e., ``quantized'') analogue becomes characterized
by a mathematically well founded
effective eigenvalue repulsion.

\begin{figure}[h]                    
\begin{center}                         
\epsfig{file=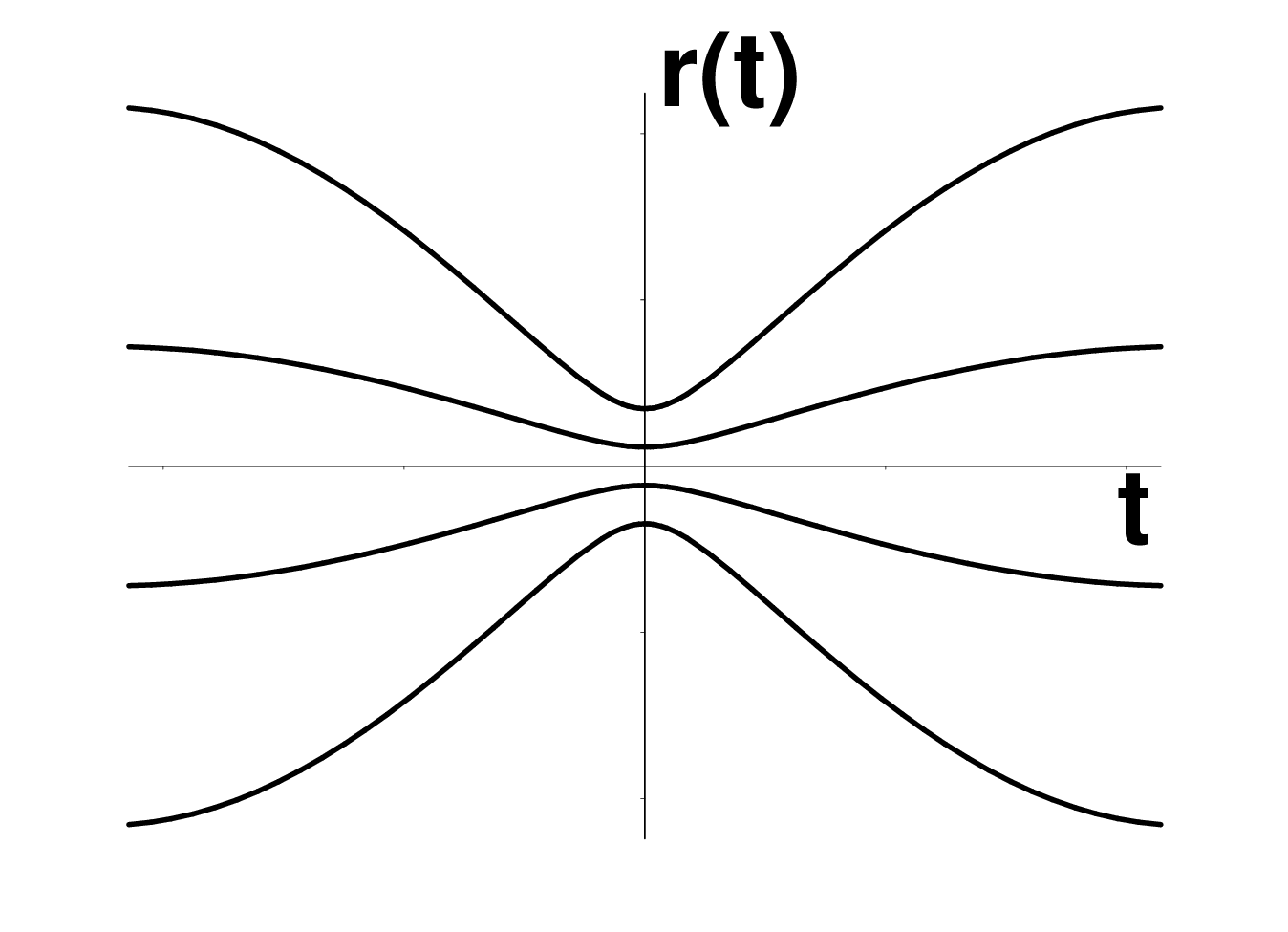,angle=270,width=0.35\textwidth}
\end{center}    
\caption{The ``avoided crossing'' behavior of eigenvalues
of an $N$ by $N$ matrix
with $N=4$
simulates the Big-Bounce regularization
of the classical Big Bang singularity at $t=0$ after quantization.
The underlying
matrix
representing a generic observable
must be Hermitian but, otherwise, it can be arbitrary.\label{globe}}
\end{figure}

In the specific context of cosmology
(which, as we emphasized, motivated our present study), the regularized
Big-Bounce phenomenon as introduced in the LQG cosmology
can be perceived as
an immediate consequence of
the assumption of the Hermiticity of the matrix.
A generic illustration of such a situation
is presented here in Figure \ref{globe}.
In its light, our present key message will be that
although
the emergence of the eigenvalue repulsion
is generic,
its mathematical foundation depends and
relies upon several tacit assumptions
which need not be satisfied in general.

\subsection{A broader theoretical framework}

\begin{figure}[h]                    
\begin{center}                         
\epsfig{file=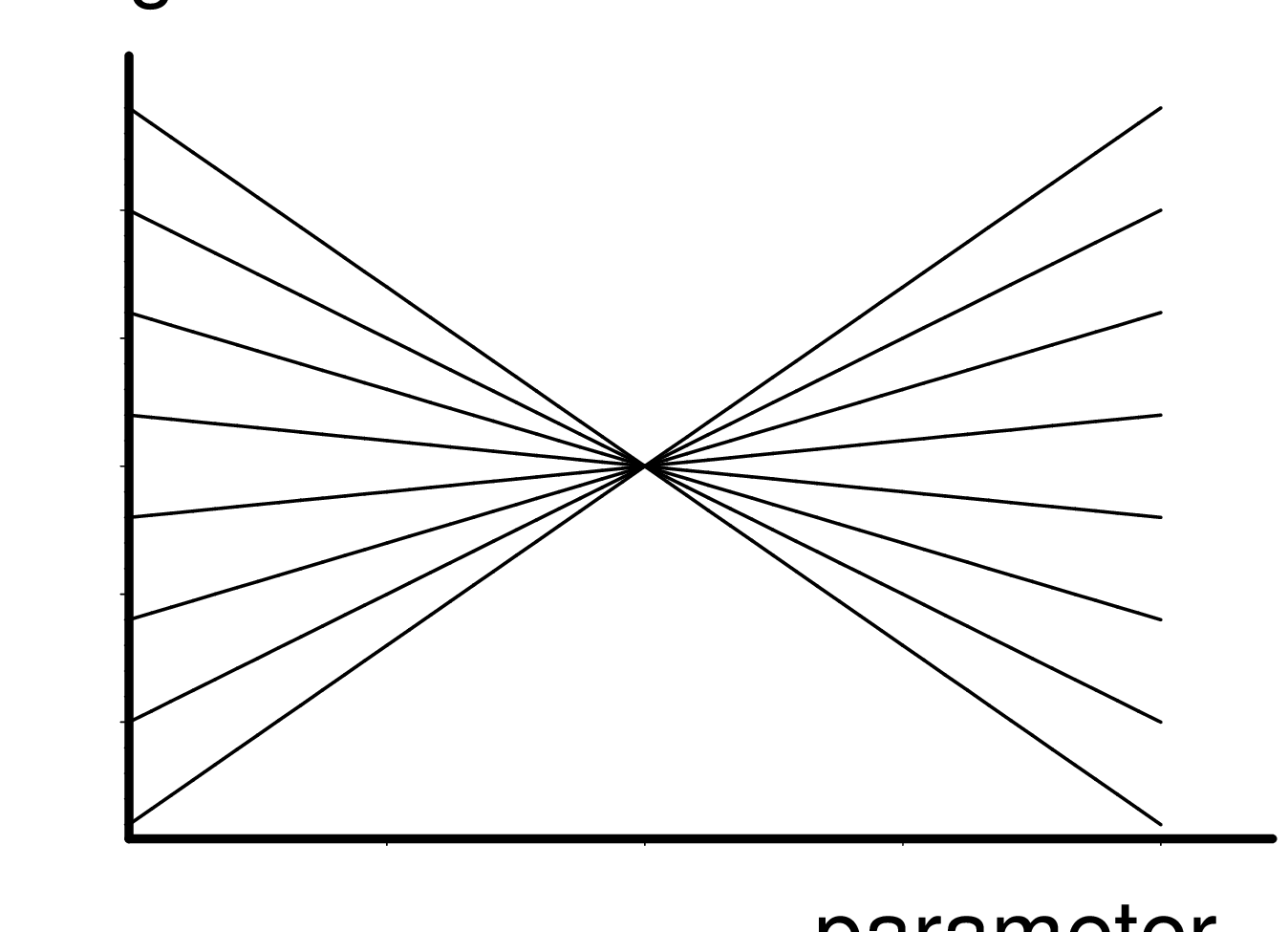,angle=270,width=0.35\textwidth}
\end{center}    
\caption{The
possibility of a
complete degeneracy
of the spectrum, i.e., of
an ``unavoided crossing'' of all
of the (real or complex) eigenvalues.
The
matrix
(here, with $N=8$)
must be
non-Hermitian (cf. Eq.~(\ref{txtc})).\label{gloja}}
\end{figure}

Once you properly weaken the assumptions,
the
statement of existence or non-existence of the singularities
after quantization
becomes an open question again.
Admitting {\em both\,} the eligible
Big Bang and Big Bounce
quantum unitary-evolution scenarios in cosmology, etc.
This
means that the overall eigenvalue-avoiding pattern
as sampled by Figure \ref{globe}
can be replaced, in principle
at least,  by
an alternative unitary-evolution scenario
admitting the eigenvalue crossing
as displayed, schematically, in Figure \ref{gloja}.


From the point of view of mathematical analysis,
Figure \ref{gloja} offers just an insufficient,
schematic and
intuitive support of our present idea
that
a strictly mathematical
phenomenon of the degeneracy of eigenvalues
could also serve
as a candidate for the description of various singular,
unsmeared quantum singularities in physics.

The gap can be found filled by the Kato's
comprehensive monograph \cite{Kato}
in which the author
proposed to call the similar degeneracies
(i.e., more precisely, the corresponding critical parameters)
``exceptional points'' (EP).
This means that, say,
in the context of quantum cosmology
one could contemplate the
EP instants
as very natural
candidates for
quantum analogues of the
singularities in classical models \cite{St,return,[2]}.

Historically, the first
implementations of the latter idea
date back to the descriptions of
non-unitary systems,
to the developments of the
theory of resonances
and to the purely  pragmatic
use of the
traditional non-Hermitian effective Hamiltonians with complex spectra, say, in the
atomic or nuclear physics \cite{Feshbach}.
In all of those contexts
one decides to relax the constraint of unitarity.
In the resulting, strongly
generalized theoretical framework
one
admits that the
toy-model operator matrices (i.e.,
in typical phenomenological applications \cite{Nimrod},
the instantaneous-complex-energy-representing
``Hamiltonians'' with $N=\infty$)
are non-Hermitian,
  \be
   R_{(energy)}^{(\infty)}(t)=H(t)
  \neq \ H^\dagger(t)
  \,.
  \label{txtcb}
  \ee
Once we admit that our formal parameter $t$ gets also complex,
the sets of the (in general, complex
but still discrete) eigenvalues $E_n(t) \in \mathbb{C}$
of $H(t)$ acquire the freedom of
merging
at the Kato's
exceptional point
values of the parameter,
 \be
 \lim_{t\to t^{(EP)}}(E_{n_1}(t)-E_{n_2}(t))=0\,.
\label{pairwb}
 \ee
It is, perhaps, worth adding that the
special, strictly pairwise (i.e., EP=EP2)
complex mergers
of Eq.~(\ref{pairwb})
can be found
in the applied perturbation theory
where
the dimension of the Hilbert space is usually infinite,
$N=\infty$,
and where they are widely known under the nickname of ``Bender-Wu
singularities'' \cite{Wu2,Wu}.

\section{Towards the survival of singularities after quantization}

The discovery of the possibility of the EP2-mediated mergers
of eigenvalues
of non-Hermitian operators  is still rather far
from
the construction of models of the Big-Bang-like quantum singularities.
Firstly, it would be necessary to guarantee that the spectrum
as
depicted, schematically, in Figure \ref{gloja}
is observable (i.e., real).
Obviously, a clarification
(or at least a partial clarification)
of the emergent related
mathematical subtleties
still is a comparatively nontrivial task.

\subsection{Unitary systems}

The essence and background of our present considerations is that
even for the strictly unitary quantum systems
(sampled above, hypothetically, by the Universe after Big Bang)
the conventional Hermiticity assumption (\ref{txtb})
can be, in a way partially paralleling Eq.~(\ref{txtcb}), weakened
to read
 \be
  R_{}^{(N)}(t)=R_{(generalized)}^{(N)}(t)\
  \neq \ \left [R_{}^{(N)}(t)\right ]^\dagger \ \
  \ \ {\rm in} \ \ \ \ {\cal H}_{(mathematical)}
\ \neq \ {\cal H}_{(physical)}
  \,.
  \label{txtc}
  \ee
One only has to add a complementary quasi-Hermiticity  \cite{Dieudonne,Geyer}
requirement
 \be
 R^\dagger(t)\,\Theta(t)=\Theta(t)\,R(t)
\label{quhe}
 \ee
where an {\it ad hoc\,}
operator $\Theta(t)$
has to
represent a correct physical inner-product
metric \cite{Geyer,timedep,SIGMA,dor,Fring} (here
we dropped the superscript
$^{(N)}$ as redundant).
With the same $\Theta(t)$
the constraint (\ref{quhe}) itself
is then assumed to be
imposed upon every
admissible candidate $R(t)$ for observable
in the underlying unitary quantum system.

An updated explanation and description of the
theory
can be found in the comprehensive
review \cite{ali}. Incidentally,
the author advocated there his opinion that
the quasi-Hermiticity
constraint (\ref{quhe})
should rather be given a more explanatory name of
$\Theta(t)-$pseudo-Hermiticity.
Such a more detailed specification may prove illuminating because
for a fixed, preselected operator
$R(t)$ the obligatory construction or choice of the
physical Hilbert-space metric $\Theta(t)$
is well known to be ambiguous \cite{SI,SIGMAdva}.
Naturally, such an
ambiguity may lead, in principle at least, to the
mutually inequivalent
implementations
and predictions of the theory \cite{Geyer,book,lotor}.

For the time being let us only add that
the use of postulate (\ref{quhe})
offers, first of all,
a straightforward resolution of the apparent
eigenvalue-repulsion
puzzle.
In applications,
it enables us to
separate clearly the choice of the set of observables $R(t)$
(with real spectra)
from the choice of the correct inner-product metric $\Theta(t)$.
This opens the way towards
the unitary-system analogue
of Eq.~(\ref{pairwb})
 \be
 \lim_{t\to t^{(EP)}}(r_{n_1}(t)-r_{n_2}(t))=0\,
\label{pairw}
 \ee
in which
one can still follow the methodical guidance by Figure \ref{gloja}.
In the new setting one
also becomes more strongly motivated to recall the Big Bang example
and to
require that the mergers (\ref{pairw})
involve, simultaneously,
all of the pairs of the
quantum numbers $n_1$ and $n_2$.
In such a case,
all of the
eigenvalues will have to degenerate to the same
exceptional point limit
of maximal order (EPN, cf., e.g., \cite{passage}).


\subsection{Schematic toy-model with EP = EPN}

In the light of Figure \ref{gloja}
an ultimate aim of
our present considerations
can be formulated as a proposal of a
family of models
in which the Hilbert-space dimension
is finite, $N<\infty$,
the spectra of the observables are real
and, last but not least, the
exceptional-point degeneracies
are
``maximal'', EP=EPN.

Naturally, the
guarantee of the spectral reality (i.e.,
of the observability of the system's characteristics)
is one of the most difficult tasks.
In essence,
one must accept assumptions
which immediately lead to a fairly deep
innovation of the formalism of quantum theory.
Forming one of the first steps
towards the above-mentioned
``prospective synthesis'' in a consistent future
quantization of gravity \cite{Ju}.

\begin{figure}[h]                     
\begin{center}                         
\epsfig{file=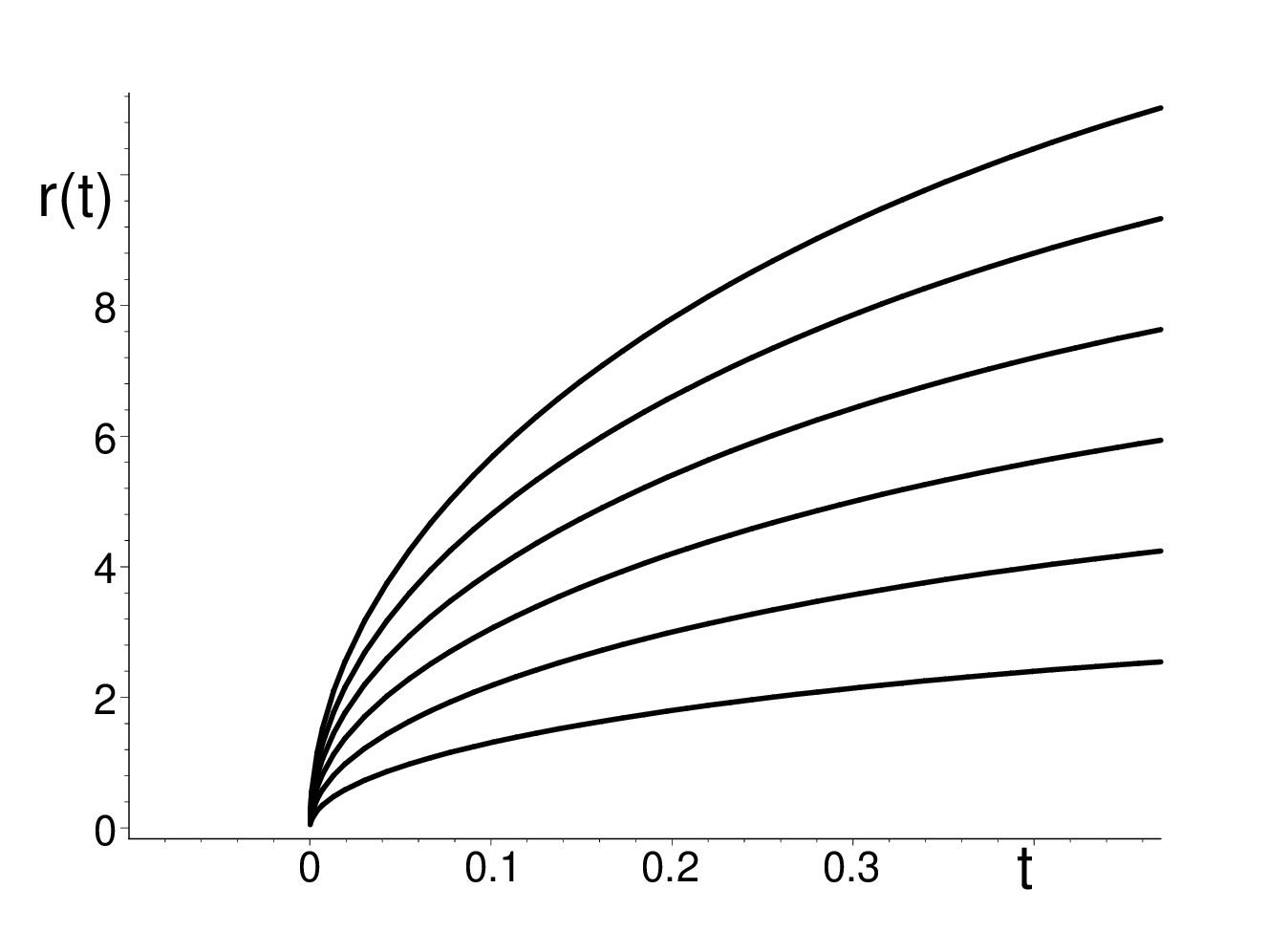,angle=270,width=0.35\textwidth}
\end{center}                         
\vspace{-2mm}\caption{Time-dependence and reality of spectrum
of the toy-model matrix of Eq.~(\ref{radaham}).
 \label{figone}}
\end{figure}

In our present paper,
the feasibility of the task will be now demonstrated using a
suitable tridiagonal-matrix toy model.
For an explicit explanation of
our present main message let us
recall the
tridiagonal-matrix
family of the $N-$level
quasi-Hermitian
observables
as proposed in  \cite{axioms}
(cf. also \cite{maximal,tridiagonal,FM}).
For the sake of simplicity
let us only pick up the
special case of this model with $N=6$.
Then,  matrix
 \be
 R_{(EPN)}^{(6)}({t}) =
 \left [\begin {array}{cccccc}
  -5+{\sigma^{}(t)}&\sqrt{5}\,{\tau^{}(t)}   &0  &0 &0  &0\\
  -\sqrt{5}\,{\tau^{}(t)}   &-3+{\sigma^{}(t)}&2\,\sqrt{2}\,{\tau^{}(t)}   &0  &0 &0\\
 0&-2\,\sqrt{2}\,{\tau^{}(t)}&-1+{\sigma^{}(t)}   &3\,{\tau^{}(t)}  &0 &0 \\
  0& 0&-3\,{\tau^{}(t)}  &1+{\sigma^{}(t)} &2\,\sqrt{2}\,{\tau^{}(t)}&0\\
  0& 0&0&-2\,\sqrt{2}\,{\tau^{}(t)}&3+{\sigma^{}(t)}&\sqrt{5}\,{\tau^{}(t)}\\
  0&0& 0&0&-\sqrt{5}\,{\tau^{}(t)}&5+{\sigma^{}(t)}
 \end {array}\right ]\,
 \label{radaham}
 \ee
with abbreviations $\tau^{}(t)=1-t\,$ and
$\sigma^{}(t)=8\,\sqrt{1-[\tau^{}(t)]^2}$
yields
the
spectrum $\{r_n(t)\}$
as displayed in Figure \ref{figone}.

%
%
%
%
%
%
%
%

In the picture we see that the
spectrum is absent (or, more precisely,
it is complex and remains unobservable)
at the negative, ``pre-Big-Bang'' times $t<0$
while it
is real and positive at $t>0$, i.e., after the fully
singular quantum Big Bang instant of degeneracy $t=t^{EPN)}=0$.
These are the reasons why we were able to interpret, in \cite{axioms},
matrices $R_{(EPN)}^{(N)}(t)$,
at every Hilbert-space dimension $N<\infty$,
as the operators
representing
radius of a schematic quantized Universe or, if one wishes,
of the ``Multiverse'' admitting a consequently singular Big Bang.

The main technical obstruction encountered
during any other, similar, phenomenologically oriented
construction of a unitary
and, perhaps, more realistic
quantum model with
singularities
is connected with the mathematical fact that
during the limiting transitions $t \to t^{(EP)}$ or $t \to t^{(EPN)}$,
all of the operators
$\Theta(t)$ guaranteeing the
quasi-Hermiticity of $R_{(EP)}^{(N)}(t)$ or $R_{(EPN)}^{(N)}(t)$
cease to be invertible. This is an important technical problem, discussed
in the rest of our present note.

\section{Emergence and localization of exceptional points
in a solvable toy model}

The situation becomes particularly challenging
when our study of a unitary quantum system
has to be based on the computer-assisted
calculations
using a suitable finite-precision arithmetics.
In such a case even the most standard and most needed numerical operations
(like, typically, the diagonalization of $R^{(N)}(t)$)
cease to be
stable in a not too small vicinity of $t^{(EP)}$.
Then, even the approximate, numerical forms of $\Theta(t)$
cease to be
eligible as acceptable
physical inner product metrics.

The problem becomes particularly serious
when we want to study the existence
and properties of EP singularities.
In such a case, indeed,
the influence of the rounding errors increases very
quickly with the decrease of the distance of $t$ from $t^{(EP)}$
(see this effect discussed in \cite{[3],[5],[12]} and
sampled in Table Nr.~1 of Ref.~\cite{EPnum}).

For our present methodical purposes we found
an elegant resolution of the problem in the choice of the
class of
certain boundary-controlled
models $R_{(bc)}^{(N)}(t)$ which appeared tractable non-numerically.

\subsection{Toy model with boundary-controlled dynamics}

For illustration purposes let us now pick up the following
time-dependent and manifestly non-Hermitian
$N$ by $N$ matrix
 \be
 R=R_{(bc)}^{(N)}(t)=
 \left[ \begin {array}{ccccc}
  2-z(t)&-1&0
 &\ldots&0
 \\
 \noalign{\medskip}-1&2&-1&\ddots&\vdots
 \\
 \noalign{\medskip}0&-1&\ddots&\ddots
 &0
 \\
 \noalign{\medskip}\vdots&\ddots&\ddots&2&-1
 \\
 \noalign{\medskip}0&\ldots&0&-1&2- z^*(t)
 \end {array} \right]\,.
 \label{Ka8t}
 \ee
The complex function of time
$z=z(t) \in \mathbb{C}$
found its origin in the so called Robin boundary conditions
as introduced in \cite{Robin}
where it has been defined, in terms
of the two real parameters $\alpha$ and $\beta$,
using formula
$z(t)=1/(1 -
\beta(t)\,h -i\alpha(t)\,h)$
where $h$ denotes an inessential real constant.

\subsection{Special case with real
rescaled time $r(t)$ in $\,z(t)={\rm i}\,\sqrt{1-r^2(t)}$}

 \noindent
In \cite{Robin} it has been proved that
in a certain interval of time
(i.e., in a certain non-empty (complex) domain ${\cal D}$ of $z(t)$),
matrix~(\ref{Ka8t})
defines an observable
(i.e., its spectrum is real).
For illustration purposes let us use, therefore,  $N=6$ and evaluate
the secular polynomial 
 $$
 \det (R_{(bc)}^{(6)}-E)
=
  {E}^{6}-12\,{E}^{5}+ \left( 56-{r}^{2} \right)E^{4}
  + \left( -128+8\,{r}^{2} \right) {E}^{3}+
$$
\be
+
 \left( 147-21\,{
 r}^{2} \right) {E}^{2}+ \left( -76+20\,{r}^{2} \right) E
 +12-5\,{r}^{2}\,.
 \ee
As long as it is just a
linear function of $r^2$,
we can extract and write down
the explicit two-branched
``Sturmian-coupling'' function $r$ of $E$,
 \be
r=r_\pm(E)=
  \pm \sqrt {
 \frac
   { {E}^{4}-8\,
{E}^{3}+ 20\,{E}^{2}-16\,E+3 }
     {{E}^{4}-8\,{E}^{3}+21\,{E}^{2}-20\,E+5 }
 }
\left( E-2 \right) \, \label{17}
 \ee
where the symbol $E=E_n(t)$ (i.e., eigenvalue) may but need not
denote the energy.

This result can easily be extended to any finite
Hilbert-space dimension $N$.
After we complement formulae like (\ref{17}) by
their respective graphical representations, it
becomes desirable and possible to prove and to conclude that
the merger of the two levels in the middle of the
spectrum really represents the EP singularity (also
known as ``non-Hermitian degeneracy'' \cite{Berry})
at $r=0$ and at any even dimension $N$.

\subsection{Shifted complex $z(t)=y+{\rm i}\,\sqrt{1-r^2(t)}$}

\begin{figure}[h]
\begin{center}
\epsfig{file=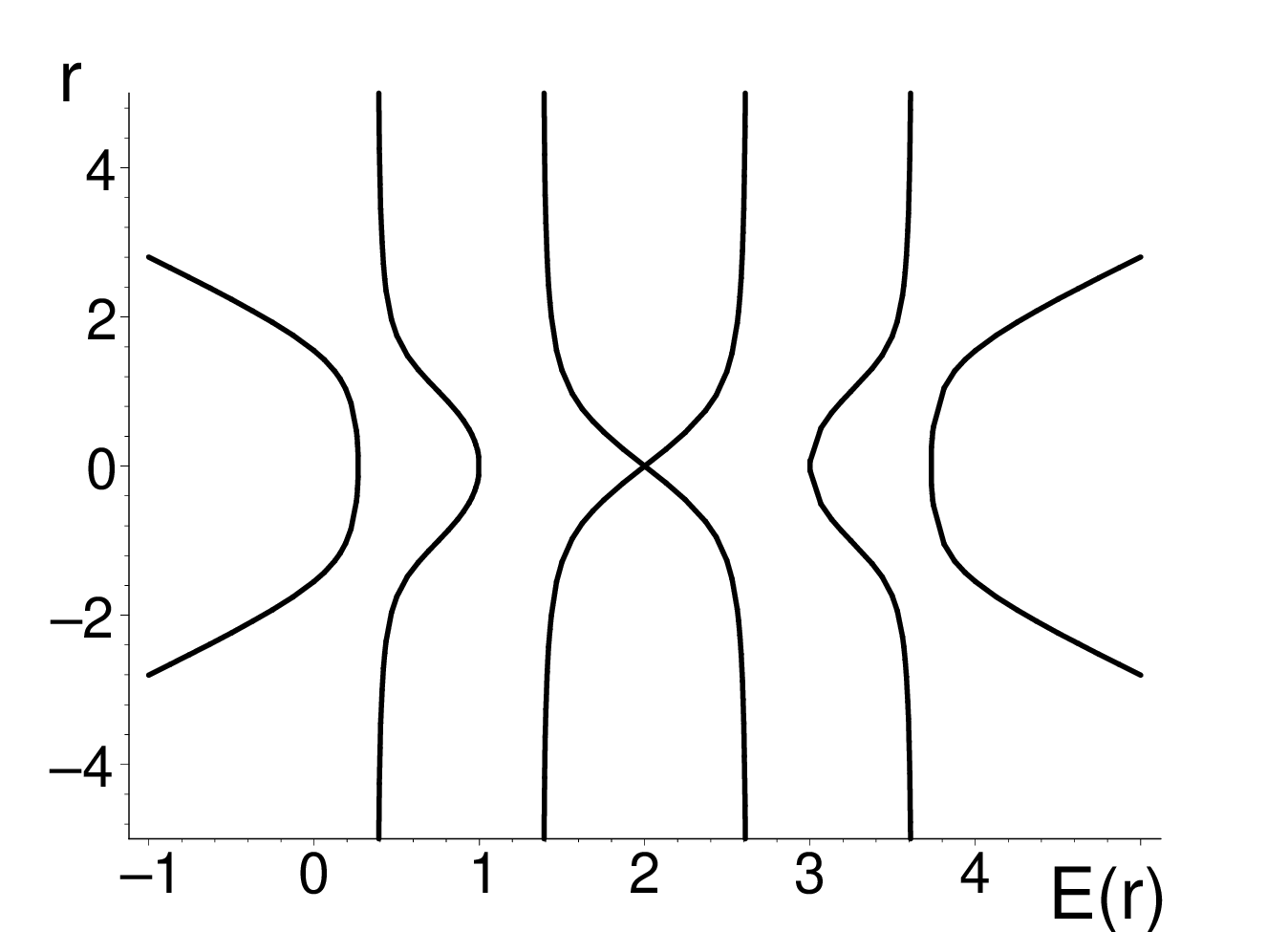,angle=270,width=0.3\textwidth}
\end{center}
\vspace{-2mm}\caption{Sturmian 
spectrum 
of matrix (\ref{Ka8t}) at $y=0$, real $r$ and $N=6$.
 \label{ufigone}}
\end{figure}

One of the slightly puzzling features of our above-outlined
toy model
is that
the spectrum only exhibits the central EP2 singularity
at the even $N$ (cf. Figure \ref{ufigone} at $N=6$).
Whenever one chooses an odd $N$ (i.e., say, $N=5$),
one reveals that the central X-shaped structure
of Figure \ref{ufigone} becomes replaced by a single
vertical straight line,
without any other really significant qualitative changes in the
picture.

This was the reason why we turned attention to
a slightly more flexible set of
complex
couplings $z(t)=y+{\rm i}\,\sqrt{1-r^2(t)}$
containing a new real and time-independent parameter $y$.
Via a detailed analysis (where we kept using $N=5$)
we revealed that the change of parity of $N$ does not really imply
the disappearance of the singularity.
In numerical tests we choose
several small and
negative values of $y< 0$ and found
that the whole spectrum $\{E_n(r)\}$ (with $n=0,1,2,3,4$)
moved,
not too surprisingly, downwards.

Purely numerically we also managed to demonstrate that
the shift is slightly
$r(t)-$dependent.
This led to the localization of
an EP2-overlap (\ref{pairwb}) with $n_1=1$ and $n_2=2$
at a non-vanishing critical value of $y=y_{1,2} \approx -0. 196$.

\begin{figure}[h]                     
\begin{center}                         
\epsfig{file=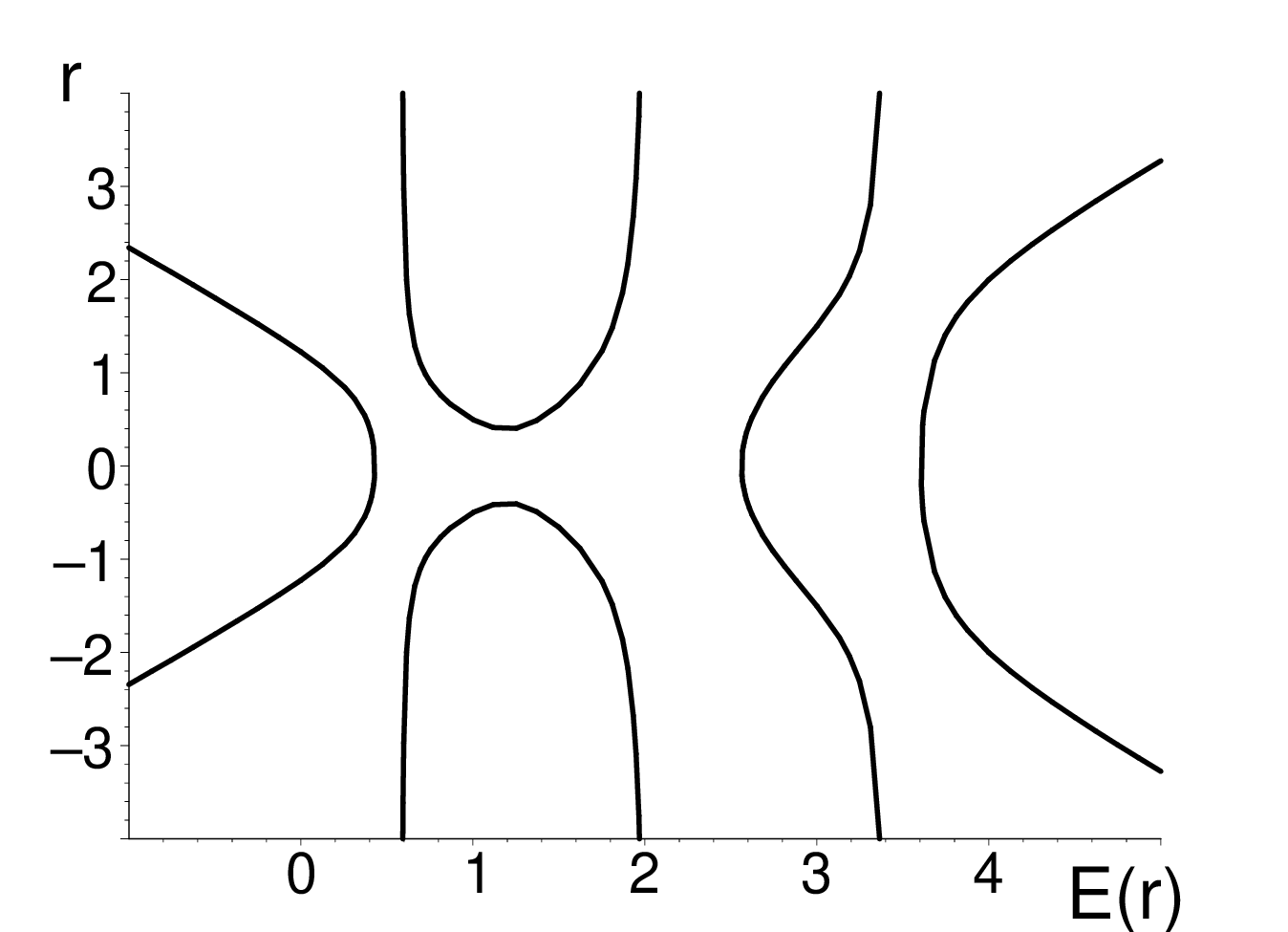,angle=270,width=0.3\textwidth}
\end{center}                         
\vspace{-2mm}\caption{Sturmian spectrum
of our toy-model matrix $R_{(bc)}^{(5)}(t)$
at $y=-0.5$.
 \label{sfigone}}
\end{figure}

After the variable $y$ crossed the latter value.
the spectrum ceased to be real.
In the corresponding interval of $y$s
the Sturmian representation of the
surviving real part of the spectrum
can be found
illustrated
by Figure \ref{sfigone}.


The latter picture indicates that
in a certain vicinity of $y=-0.5$
one obtains Im $E_1(r) \neq 0$ and Im $E_2(r) \neq 0$
in a certain not too large interval
of $r\in (-r_{critical}(y),r_{critical}(y))$.

\begin{figure}[h]                     
\begin{center}                         
\epsfig{file=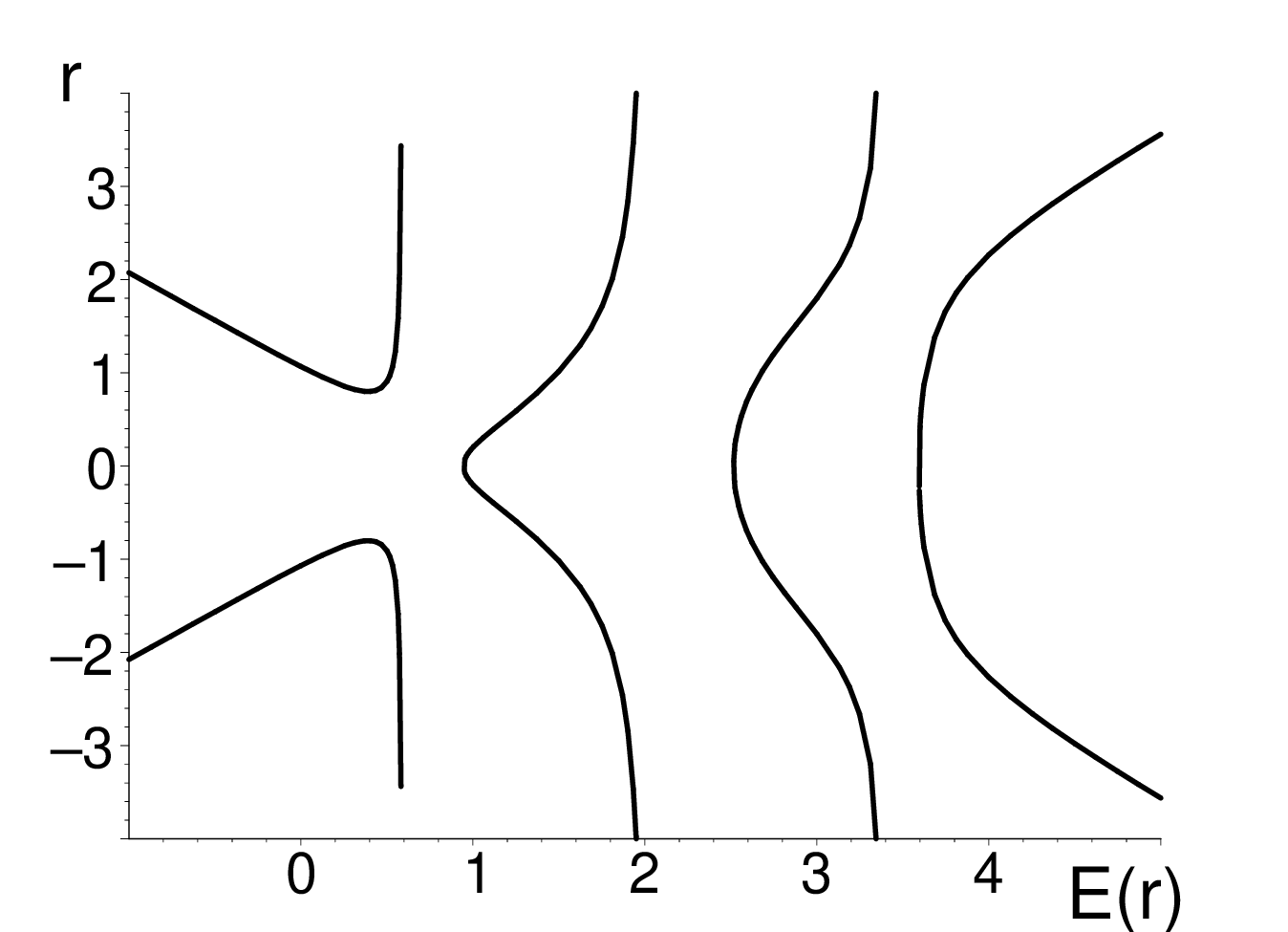,angle=270,width=0.3\textwidth}
\end{center}                         
\vspace{-2mm}\caption{Sturmian spectrum
of our toy-model matrix $R_{(bc)}^{(5)}(t)$
at $y(t)=-0.8$.
 \label{rfigone}}
\end{figure}

After a further decrease of
the value of $y$
the full reality of spectrum
is never recovered.
Although the level $E_2(r)$ re-acquires its
reality at all $r$ at
another critical value of $y \approx -0.7071$,
this is not an EP
singularity because Sturmian function $r=r(E)$
has just a simple pole there.
The loss of observability
becomes transferred to
the ground state, with
the new
structure of the Sturmian spectrum
sampled here by our last Figure~\ref{rfigone}.

\section{Summary}

Besides the obligatory quasi-Hermiticity
(\ref{quhe}) imposed upon any relevant observable $R(t)$
the description of the
quasi-Hermitian quantum theory would remain incomplete without
a clarification of the relationship between the two mutually non-equivalent
Hilbert spaces
which emerged in Eq.~(\ref{txtc}),
viz., between
${\cal H}_{(mathematical)}$ and ${\cal H}_{(physical)}$.
In fact, the only subtlety of this clarification lies in
the necessity of using an appropriate notation.

%

The price to pay lies in the necessity of a disentanglement of a
potentially misleading terminology used in the current literature.

\begin{description}

\item[(a)]
In a way already used, without warning, in Eq.~(\ref{txtc}),
the superscript $^\dagger$ is {\em always\,} used to indicate the
{\em unphysical\,}
Hermitian conjugation in ${\cal H}_{(mathematical)}$.
As a consequence,

\begin{description}

\item[(i)]
our operators of observables (which are, naturally,
self-adjoint in ${\cal H}_{(physical)}$) seem to be
non-Hermitian
in the unphysical and phenomenologically irrelevant Hilbert
space ${\cal H}_{(mathematical)}$.
So, we should not call them
non-Hermitian. But we do.

\item[(ii)]
Much less frequently we call our observables, correctly,
quasi-Hermitian, having in mind their
restriction by the metric-dependent constraint (\ref{quhe})
in ${\cal H}_{(mathematical)}$.

\end{description}

\item[(b)]
Tacitly, all of the standard Dirac's bra-ket conventions are
assumed to be used
in the ``non-textbook Hilbert space'' ${\cal H}_{(mathematical)}$
rather than in its
``textbook space'' avatar
${\cal H}_{(physical)}$.
The exclusive correct probabilistic interpretation
of the theory in the latter space
is only indirectly kept in mind.

\begin{description}

\item[(iii)]
No problems emerge with the use of kets
$|\psi\kt$
since, by definition, they denote the same, identical
element $\psi$ of both of the two vector spaces
${\cal H}_{(mathematical)}$ and ${\cal H}_{(physical)}$ \cite{ali}.

\item[(iv)]
A minor nontrivial amendment of the notation conventions
is only recommended (cf. \cite{SIGMA,jupi}) in connection
with the introduction of a very useful
``ketket'' abbreviation $|\psi\kkt :=\Theta(t)\,|\psi\kt$.

\end{description}

\end{description}

As long as we always work in the unique and preselected
Hilbert space ${\cal H}_{(mathematical)}$,
no problems arise also due to our use of the respective
(i.e., ``bra'' and ``brabra'') duals. Thus, beyond the unphysical dual
$\br \psi| \ \in \ [{\cal H}_{(mathematical)}]'$
we need
the physical-space linear functional
$\bbr \psi| \ \in \
[{\cal H}_{(physical)}]'$
represented, without problems, by the
formal conjugation of $|\psi\kkt$ in
our unique representation Hilbert space
${\cal H}_{(mathematical)}$.
We can conclude that
the essence of the pedagogical difficulty lies in a
slightly counterintuitive nature of the theory
in which the well known and physical Hilbert space
${\cal H}_{(physical)}$
is practically never explicitly recalled
in applications.
Always, we only refer to its representation
in ${\cal H}_{(mathematical)}$
using the so called ``inner-product metric'' operator $\Theta(t)$.
Thus, we build the models and predictions in
the space ${\cal H}_{(mathematical)}:={\cal H}$
which is
manifestly unphysical but 
maximally 
user-friendly.


\newpage

\begin{thebibliography}{00}


\bibitem{Rovelli}
Rovelli C 2004 {\it Quantum Gravity} (Cambridge: Cambridge University Press)

\bibitem{Thiemann}
Thiemann T 2007 {\it Introduction to Modern Canonical Quantum General
Relativity} (Cambridge: Cambridge University Press)

\bibitem{Bounce}
Bojowald M 2001  {\it Phys. Rev. Lett.} {\bf 86} 5227--5230



    \bibitem{piech}
Ashtekar A, Pawlowski T and Singh P 2006
{\it Phys. Rev.} D {\bf 74}  084003



\bibitem{Messiah}
Messiah  A 1962 {\it Quantum Mechanics} (Amsterdam: North Holland)
%
%




\bibitem{Kato}
Kato T 1966 {\it Perturbation Theory for Linear Operators}
(Berlin: Springer-Verlag)

%



\bibitem{St}
Heiss W D, M\"{u}ller M and Rotter I 1998
 {\it Phys. Rev.} E {\bf 58} 2894

\bibitem{return}
Znojil M 2007
 {\it Phys. Lett.} B  {\bf 647} 225 -
 230
%


\bibitem{[2]}
Znojil M  
2019
 {\it Phys. Rev.} A  {\bf 100} 032124




\bibitem{Feshbach}
Feshbach H 1958  {\it Ann. Phys. (NY)} {\bf 5} 357--390
%
%
%




\bibitem{Nimrod}
Moiseyev N 2011 {\it Non-Hermitian Quantum Mechanics} (Cambridge:
CUP)



\bibitem{Wu2}
Bender C M and
Wu T T 1969
 {\it Phys. Rev.}  {\bf 184} 1231

\bibitem{Wu}
%
Alvarez G 1995
 {\it J. Phys. A: Math. Gen.}  {\bf 28} 4589--4598

\bibitem{Dieudonne}
Dieudonn\'{e} J 1961
{\it Proc. Int. Symp. Lin. Spaces} (Oxford: Pergamon)  pp 115--122
%
%



\bibitem{Geyer}
Scholtz F G, Geyer H B and Hahne F J W 1992 {\it Ann. Phys. (NY)}
{\bf 213} 74--101

\bibitem{timedep}
Znojil M 2008
 {\it Phys. Rev. D} {\bf 78}
 085003
%
%


\bibitem{SIGMA}
Znojil M 2009  {\it Symm. Integ. Geom. Methods and Applications}
{\bf 5} 001
%
%
%


\bibitem{dor}
Brody  D C 2014
 {\it J. Phys. A : Math. Theor.} {\bf 47}
  035305


\bibitem{Fring}
Fring A and Moussa M H Y 2016
 {\it Phys. Rev. A} {\bf 93}
  042114

\bibitem{ali}
Mostafazadeh A 2010
 {\it Int. J. Geom. Meth. Mod. Phys.} {\bf 7} 1191--1306
%
%


\bibitem{SI}
Znojil M and Geyer H B
2006   {\it Phys. Lett. B} {\bf 640} 52 - 56


\bibitem{SIGMAdva}
Znojil M 2008  {\it Symm. Integ. Geom. Meth. Appl. SIGMA} {\bf 4} 001
%

\bibitem{book}
Bagarello F, Gazeau J P, Szafraniec F H and Znojil M 2015 {\it
%
Non-Selfadjoint Operators in Quantum Physics: Mathematical Aspects}
 ed F Bagarello, J-P Gazeau et al
  (Hoboken, NJ: John Wiley \& Sons)

%

\bibitem{lotor}
Krej\v{c}i\v{r}\'{\i}k D, Lotoreichik V and Znojil M 2018
 {\it Proc. Roy. Soc. A: Math. Phys. Eng. Sci.}  {\bf 474} 20180264


\bibitem{passage}
Znojil M 2020
 {\it Proc. Roy. Soc.} A  {\bf 476} 20190831
%


\bibitem{Ju}
Ju  C Y,
 Miranowicz A, Chen Y N, Chen  G Y and Nori F 2024
{\it QUANTUM} {\bf 8}
1 - 20

\bibitem{axioms}
Znojil M 2023  {\it Axioms} {\bf 12} 644
%



\bibitem{maximal}
Znojil M 2007
 {\it J. Phys. A: Math. Theor.}  {\bf 40} 4863--4875

\bibitem{tridiagonal}
Znojil M 2007
 {\it J. Phys. A: Math. Theor.}  {\bf 40} 13131--13148


\bibitem{FM}
Fern\'{a}ndez F M 2022
 {\it Ann. Phys. (NY)} {\bf 443} 169008



\bibitem{[3]}
Znojil M 2018 {\it Phys. Rev.} A {\bf 97} 032114
%


\bibitem{[5]}
Znojil M 2020
 {\it Entropy}  {\bf 22} 000080 

\bibitem{[12]}
Znojil M 2020
 {\it Symmetry}  {\bf 12} 1309

\bibitem{EPnum}
Znojil M 2019
 {\it Ann. Phys.} (NY) {\bf 405}  325--339
%
%
%
%



\bibitem{Robin}
Znojil M 2014  {\it J. Phys. A: Math. Theor.} {\bf 47} 435302

\bibitem{Berry}
Berry M V  2004  {\it Czech. J. Phys.} {\bf 54} 1039--1047
%

\bibitem{jupi}
Ohmori S and Takahashi J
 2022  {\it J. Math. Phys.} {\bf 63} 123503


\end{thebibliography}
\end{document}